\documentclass[prl, reprint, superscriptaddress]{revtex4-2}
\usepackage{amsmath,subcaption,graphicx, xcolor}

\begin{document}
\title{Transverse Chiral Magnetic Photocurrent\\ Induced by Linearly Polarized Light in\\ Mirror-Symmetric Weyl Semimetals}
\date{\today}
\author{Sahal Kaushik}
\email{sahal.kaushik@stonybrook.edu}
\affiliation{Department of Physics and Astronomy, Stony Brook University, Stony Brook, New York 11794-3800, USA}
\author{Dmitri E. Kharzeev}
\email{dmitri.kharzeev@stonybrook.edu}
\email{kharzeev@bnl.gov}
\affiliation{Department of Physics and Astronomy, Stony Brook University, Stony Brook, New York 11794-3800, USA}
\affiliation{Department of Physics, Brookhaven National Laboratory, Upton, New York 11973-5000, USA}
\affiliation{RIKEN-BNL Research Center, Brookhaven National Laboratory, Upton, New York 11973-5000, USA}
\author{Evan John Philip}
\email{evan.philip@stonybrook.edu}
\affiliation{Department of Physics and Astronomy, Stony Brook University, Stony Brook, New York 11794-3800, USA}

\begin{abstract}
A new class of photocurrents is predicted to occur in both type-I and type-II Weyl semimetals. Unlike the previously studied photocurrents in chiral materials, the proposed current requires neither circularly polarized light, nor an absence of symmetry with respect to a plane of reflection. 
We show that if a Weyl semimetal has a broken inversion symmetry then linearly polarized light can induce a photocurrent transverse to the direction of an applied magnetic field, in spite of the symmetry with respect to a reflection plane and the time reversal symmetry. The class of materials in which we expect this to occur is sufficiently broad and includes the transition metal monopnictides such as TaAs. The effect stems from the dynamics of Weyl chiral quasi-particles in a magnetic field, restricted by the symmetries described above; because the resulting current is transverse to the direction of magnetic field, we call it the \emph{transverse chiral magnetic photocurrent}. The magnitude of the resulting photocurrent is predicted to be significant in the THz frequency range, about $0.75\; \mathrm{\mu A}$ for type-I and $2.5\; \mathrm{\mu A}$ for type-II Weyl semimetals.
This opens the possibility to utilize the predicted transverse chiral magnetic photocurrent for sensing unpolarized THz radiation. 

\end{abstract}
\maketitle

The hallmark of Dirac and Weyl semimetals is the emergence of chiral fermionic quasiparticles; for a review, see \citet{armitage2018}. The chiral quasiparticles enable the study of quantum anomaly-induced phenomena, such as the chiral magnetic effect~\cite{2008Fukushima}, which results in a large negative magnetoresistance~\cite{2013Son,2014Burkov} observed recently in a number of Dirac~\cite{2016Li, 2015Xiong} and Weyl~\cite{Huang2015, Zhang2016, Wang2016, Arnold2016} semimetals. Dirac and Weyl fermions possess well-defined chirality $\chi = \mathrm{sgn}(\vec{s}\cdot\vec{q})$ (where $\vec{s}$ is the pseudospin, and $\vec{q}$ is the crystal momentum) that is approximately conserved because the rate of chirality flipping transitions is usually much smaller than the rate of chirality-preserving scattering. 

In Dirac materials, the left- and right-handed fermions are at the same points in the Brillouin zone, while in Weyl materials, they are at different points. The existence of Weyl cones is topologically protected, as a small perturbation cannot open a gap in the quasiparticle spectrum. Dirac cones are not topologically protected, but may be protected by crystal symmetries~\cite{Yang2014}. If a material has both inversion and time reversal symmetries, the cones of different chiralities must necessarily lie at the same points in the Brillouin zone, so Weyl materials have either broken inversion, or broken time reversal symmetries, or both. In this work, we focus on Weyl materials with broken inversion symmetry. Many Weyl semimetals, including TaAs have a symmetry with respect to at least one plane of reflection; we will refer to such materials as mirror-symmetric, and to those that lack such a symmetry as asymmetric.

Recently, a chiral photocurrent has been predicted~\cite{lee2017} and observed~\cite{ma2017} in the Weyl semimetal TaAs in response to circularly polarized light. This photocurrent results from  the combined effect of the selection rules involving chirality and the tilt of the Weyl cones. The emission of elliptically polarized THz radiation in response to circularly polarized infrared light, as a result of transitions between Weyl and non-Weyl bands, has also been observed recently in TaAs~\cite{gao2020chiral}.

Photocurrents induced by magnetic fields have been theoretically studied and experimentally observed in quantum well systems ~\cite{2005belkov} 
 The helical magnetic effect has been proposed~\cite{kharzeev2018giant} to cause colossal photocurrents parallel to an applied magnetic field in asymmetric Weyl semimetals. This effect is due to a combination of Pauli blockade and the effects of Berry curvature and magnetic field. Strong magnetic fields have been predicted to induce a magnetogyrotropic photogalvanic current in Weyl semimetals with tilted cones, due to the quantization of Landau levels~\cite{2018golub}. In mirror-symmetric Dirac and Weyl semimetals, in the presence of a magnetic field, a photocurrent has also been predicted due to chiral anomaly~\cite{kaushik2019chiral}. Circularly polarized light has also been predicted to cause a topologically protected photocurrent through the quantized circlular photogalvanic effect in asymmetric Weyl semimetals~\cite{dejuan2017}, and a photocurrent parallel to the direction of incident light due to an induced effective magnetic field~\cite{2016PhRvB..93t1202T}.

In all these previously discussed cases, one relies either on circularly polarized light or on an asymmetric crystal structure to induce the photocurrent. In this paper, we propose a new class of photocurrents that appear when the material has a broken inversion symmetry ($x,y,z \to -x, -y, -z$) but unbroken time-reversal symmetry ${\rm T}$, in the presence of a background magnetic field. Indeed, the broken inversion symmetry allows the current of the following structure:
\begin{equation}
j^i = (\sigma_B)^i_j \ B^j,
\end{equation}
where $B^j$ is the magnetic field. 
Since under inversion transformation $B^j \to B^j$ and $j^i \to - j^i$, the current is allowed by parity only if the material lacks the inversion symmetry under which $(\sigma_B)^i_j \to -(\sigma_B)^i_j$. 

In a material with a  plane of reflection symmetry, the components of current and magnetic field parallel ($\parallel$) and perpendicular ($\perp$) to the plane transform under reflection as
\begin{align*}
   \vec{j}_\perp &\to  -\vec{j}_\perp\,, \\
   \vec{j}_\parallel &\to  \phantom{-}\vec{j}_\parallel\,, \\
   \vec{B}_\perp &\to  \phantom{-}\vec{B}_\perp\,, \\
   \vec{B}_\parallel &\to  -\vec{B}_\parallel\,.
\end{align*}
Therefore, in a material with broken inversion symmetry but reflection symmetry about at least one plane, $(\sigma_B)^i_j$ cannot have a component proportional to $\delta^i_j$ but it can still have transverse components of the form $(\sigma_B)^\perp_\parallel$ and $(\sigma_B)_\perp^\parallel$.  If we  have neither inversion nor reflection symmetry, $(\sigma_B)^i_j$ is allowed to have a component proportional to $\delta^i_j$; this component corresponds to the helical magnetic effect~\cite{kharzeev2018giant}. For a related symmetry analysis, see \citet{silva2020magneticconductivity}.

Transition metal monopnictides such as TaAs are examples of mirror-symmetric Weyl semimetals; even though they have broken inversion symmetry, they have reflection symmetry about multiple planes. They have tetragonal symmetry with the space group $I 4_1 md$ (No. 109): in particular, they have fourfold rotation symmetry about the $c$ axis, reflection symmetries about the $ac$ and $bc$ planes, but no reflection symmetry about the $ab$ plane.

The photocurrents in TaAs in response to circularly polarized light discussed and observed so far~\cite{ma2017, lee2017, gao2020chiral} are necessarily of the form $\vec{j} \sim \hat{c}\times\vec{J}$, where $\vec{J}$ is the angular momentum of the incident light. This is because of the combination of fourfold rotation and reflection symmetries, as we will now explain.

We can write the photocurrent as $j^i = (\sigma_P)^i_k J^k$. Since angular momentum and magnetic field have the same transformations under reflection,  reflection symmetry imposes the same constraints on the tensors $(\sigma_P)^i_j$ and $(\sigma_B)^i_j$. Since there is reflection symmetry about the $ac$ plane, the only non-vanishing components are $(\sigma_P)^a_b$, $(\sigma_P)^b_a$, $(\sigma_P)^c_b$, and $(\sigma_P)^b_c$. Similarly, due to reflection symmetry about the $bc$ plane, the components $(\sigma_P)^c_b$, and $(\sigma_P)^b_c$ also vanish. Due to the fourfold rotation symmetry about the $c$ axis, there is only one independent component $(\sigma_P)^b_a = - (\sigma_P)^a_b \equiv \sigma_P$, and the photocurrent is therefore 
\begin{equation}
    \vec{j} = \sigma_P\; \hat{c}\times\vec{J}.
\end{equation}

Since magnetic field $\vec{B}$ is even under parity and odd under time reversal, just like  angular momentum $\vec{J}$, we can use similar symmetry analyses for photocurrents in the presence of magnetic field and photocurrents induced by circularly polarized light. Such an analysis for the photocurrent due to unpolarized light in the presence of a magnetic field yields that the current should necessarily be of the form 
\begin{equation}\label{eq:current1}
    \vec{j} = \sigma_B\; \hat{c}\times\vec{B}
\end{equation}
Unlike the helical magnetic photocurrent~\cite{kharzeev2018giant} and chiral magnetic photocurrent~\cite{kaushik2019chiral} , this photocurrent is perpendicular to the magnetic field.

\begin{figure}[htp]
  \includegraphics[]{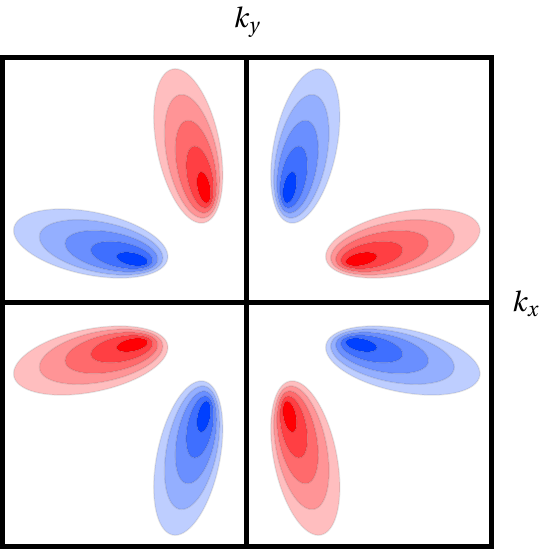}
  \caption{Weyl cones which possess both anisotropy and tilt. These Weyl cones are similar to the $W_1$ cones of the prototypical type-I Weyl semimetal used for numerical calculations in this paper. The shading represents energy and the color represents chirality.}
  \label{fig:TiltAniso}
\end{figure}

The symmetry analysis above shows that this photocurrent is not forbidden. We will now demonstrate how the anisotropy and tilt of the Weyl cones (shown in Figure~\ref{fig:TiltAniso}), combined with the chirality of quasi-particles, enable this photocurrent. Recall that the resulting current in Eq.~\eqref{eq:current1} is transverse to the direction of the applied magnetic field. This is why we call it the \emph{transverse chiral magnetic photocurrent}. The transverse chiral magnetic photocurrent is due to the interplay between the modification of velocity of chiral quasi-particles by Berry curvature and the tilt and anisotropy of the Weyl cones that prevent the cancellation of the resulting current due to the symmetries of the Brillouin zone. 

\begin{figure}[htp]
     \centering
     \begin{subfigure}[b]{0.49\linewidth}
         \centering
         \includegraphics[]{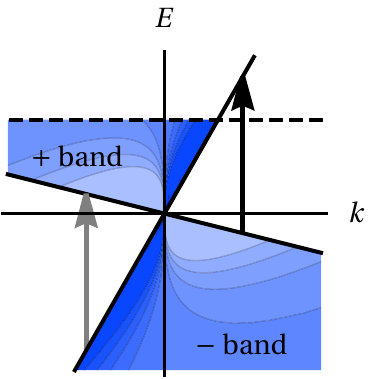}
     \end{subfigure}
     \hfill
     \begin{subfigure}[b]{0.49\linewidth}
         \centering
         \includegraphics[]{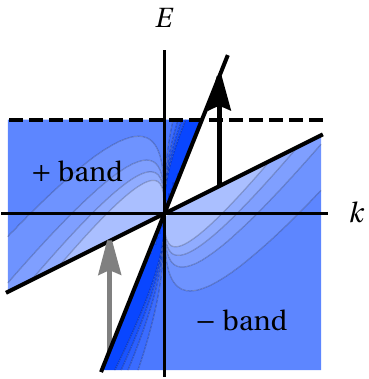}
     \end{subfigure}
        \caption{Transitions at zero temperature for type-I (left) and type-II (right) Weyl semimetals. The shading represents the effect of the Berry curvature on the density of states. The black arrow represents transitions that are allowed, and the grey arrow represents transitions forbidden by Pauli blockade at zero temperature.}
        \label{fig:WeylCones}
\end{figure}

We will now calculate the magnitude of the transverse chiral magnetic photocurrent for a prototypical Weyl semimetal with tetragonal symmetry, similar to transition metal monopnictides that include TaAs.
We assume that the crystalline structure has a plane of reflection, combination of rotation and reflection, or a combination of translation and reflection as a symmetry. The Hamiltonian of a linear Weyl cone has a general structure  
\begin{equation}\label{eq:Ham}
H = v_T^i q_i + v_a^i \sigma_a q_i,
\end{equation}
where $q_i$ is the crystal momentum of the quasi-particle and $\sigma_a$ are the Pauli matrices. For each cone, the vector $v_T^i$ characterizes its tilt. We can define the tensor $(v_W^2)^{ij} = v_a^i v_a^j$ and its inverse  $(v_W^{-2})_{ij}$ which characterize the anisotropy of the second term of the Hamiltonian in Eq.~\eqref{eq:Ham}. The chirality of the Weyl cone is defined as $\chi = \mathrm{sgn}\left( \mathrm{det}\left(v_a^i\right)\right)$. We can also define a dimensionless tilt parameter $\alpha = \sqrt{(v_W^{-2})_{ij}v_T^i v_T^j}$. In a type-I Weyl semimetal $\alpha < 1$ and in a type-II Weyl semimetal $\alpha>1$. Type-I Weyl semimetals have point-like (or in general compact) Fermi surfaces and type-II Weyl semimetals have open Fermi surfaces~\cite{soluyanov2015type}. The Weyl cones for type-I and type-II Weyl materials are shown in Figure~\ref{fig:WeylCones}.

Weyl points are monopoles of Berry curvature and their chirality is the sign of the Berry monopole charge. Because the total Berry flux through the Brillioun zone is zero, there is always an equal number of left-handed and right-handed Weyl points (Nielsen-Ninomiya theorem~\cite{1981Nielsen}). The Berry curvature depends on the eigenstates, not on the corresponding energy eigenvalues of the Hamiltonian, and is therefore independent of the tilt velocity and depends only on the untilted part of the Hamiltonian. For our Hamiltonian, the Berry curvature is
\begin{equation}
\Omega_i^\pm = \pm\chi \frac{1}{2}\frac{q_i}{(q_j (v_W^2)^{jk}q_k)^{3/2}}\sqrt{\det(v_W^2)},
\end{equation}
where the $+$ and $-$ are for the upper and lower Weyl bands (shown in Figure~\ref{fig:WeylCones}).

In the semiclassical limit, i.e. when the cyclotron frequency $\omega_c = eBv/k$ is much smaller than the inverse of the scattering time $1/\tau$ and the temperature $T$, we can ignore the Landau quantization~\cite{2015gustavo}, and model the combined effects of a static magnetic field and momentum-space Berry curvature as a modification of the phase space ~\cite{xiao2005berry,stephanov2012chiral,son2013kinetic}
\begin{equation}
\frac{d^3 q_i}{(2\pi)^3} \to (1+e\Omega_j B^j)\frac{d^3 q_i}{(2\pi)^3}
\end{equation}
and velocity is modified as
\begin{equation}
v^i \to \frac{1}{1+e\Omega_j B^j} [v^i + e(v^j\Omega_j) B^i].
\end{equation}
Since the rate of transitions $\Gamma$ depends on the number of initial states and the number of final states, the effect of phase space cancels out:
\begin{equation}
\Gamma \to \Gamma (1+e\Omega_j^+ B^j)(1+e\Omega_j^- B^j) \approx \Gamma ,
\end{equation}
and we only have to consider the modifications to velocity:
\begin{multline}\label{eq:velocity}
v^i_+ - v^i_-  \to (v^i_+ - v^i_-)  -(v^i_+ + v^i_-)(e\Omega_j^+ B^j)\\ + e(v^j_+ + v^j_-)\Omega_j^+ B^i .
\end{multline}
There is no photocurrent in the absence of $B^i$. By reflection symmetry, there can also be no term along $B^i$, so we will focus on the second term in Eq.~\eqref{eq:velocity}. Since $v^i_+ + v^i_- = 2v_T^i$, the only relevant term in Eq.~\eqref{eq:velocity} is given by
\begin{equation}\label{eq:simpl_v}
v^i_+ - v^i_-  \sim   -2v_T^i(e\Omega_j^+ B^j).
\end{equation}

\begin{widetext}
Making the assumption that one absorbed photon excites exactly one electron, the sheet current density, that is the current density integrated over the penetration depth $\int \vec{j} dz$, can be obtained by convoluting the change of velocity given by  Eq.~\eqref{eq:velocity} with the rate of transitions induced by photons:
\begin{equation}\label{eq:current}
j^i = \frac{I[1-R(\omega)]}{\hbar \omega} e  \frac{\sum_{cones}\tau \int \frac{d^3 q_j}{(2\pi)^3} \Gamma_W[f(E_-) - f(E_+)] (-2v_T^ie\Omega_k^+ B^k)\delta (E_{+} - E_{-} - \hbar \omega)} {\Gamma_{abs}(\omega) + \sum_{cones} \int \frac{d^3 q_j}{(2\pi)^3} \Gamma_W[f(E_-) - f(E_+)] \delta (E_{+} - E_{-} - \hbar \omega)} ,
\end{equation}
\end{widetext}
where $\tau$ is the relaxation time,  $f(E)$ is the Fermi distribution function, $I$ is the intensity of the incident light and $R(\omega)$ is the reflectivity. $\Gamma_{abs}(\omega)$ is the rate of absorption from mechanisms not involving chiral fermions, and $\Gamma_W$ is the rate of transitions for Weyl fermions, which is given by Fermi's golden rule $\Gamma_W \sim |\langle \psi_- | V | \psi_+ \rangle|^2$. The interaction term $V$ can be obtained by Peierls substitution, by replacing $q_i$ by $q_i + eA_i$ in the Hamiltonian as 
\begin{equation}
V  = H(q_i+eA_i) - H(q_i) \sim v_T^i \epsilon_i + v_a^i \sigma_a \epsilon_i,
\end{equation}
where $\epsilon_i$ is the polarization vector. For linearly polarized light, the transition rate $\Gamma$ is of the form
\begin{equation}
\Gamma_W \sim \epsilon_i (v_W^2)^{ij} \epsilon_j - \frac{(\epsilon_i (v_W^2)^{ij} q_j)^2}{q_i (v_W^2)^{kl} q_k}.
\end{equation}

\begin{figure*}[htp]
  \includegraphics[]{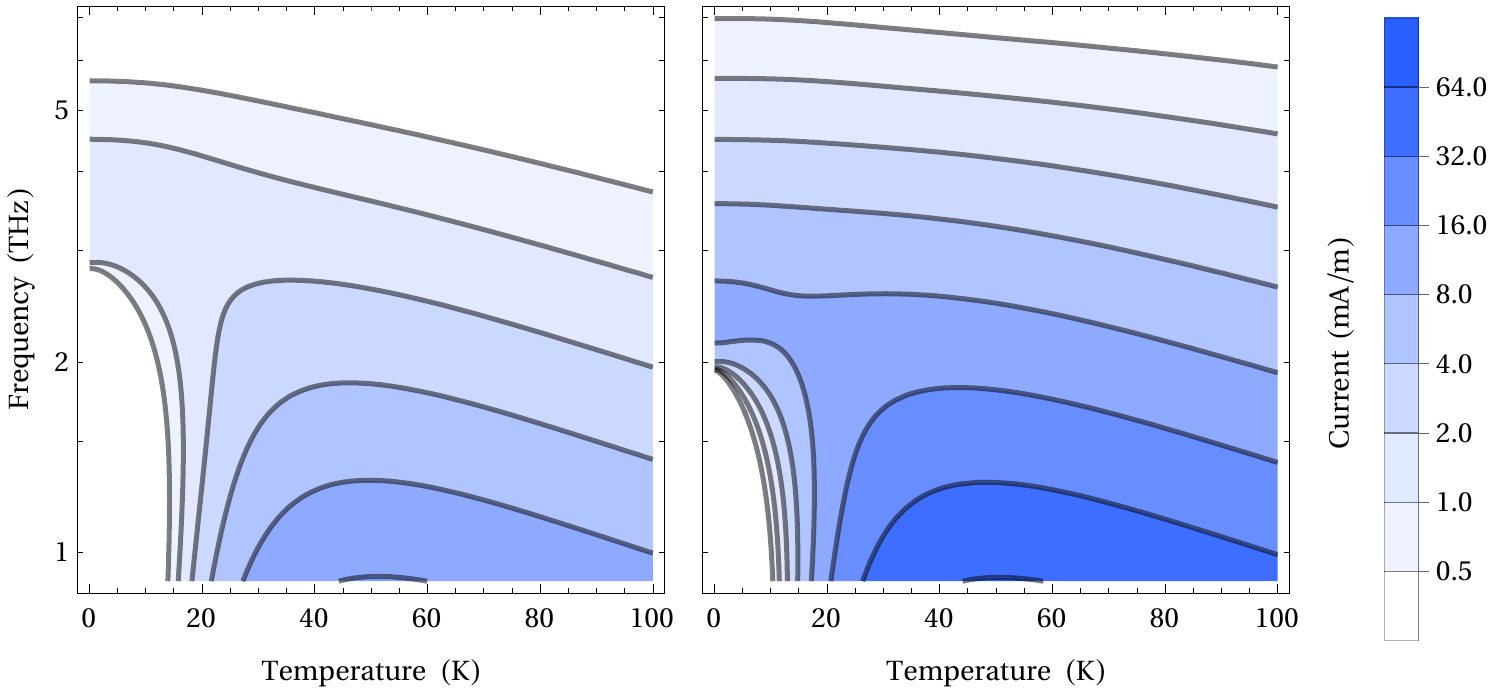}
  \caption{Photocurrent density of the prototypical type-I (left) and type-II (right) Weyl semimetal.}
  \label{fig:ResultPlot}
\end{figure*}
In the numerical calculations, we consider absorption only due to excitation of chiral fermions. Under this assumption, we can ignore the constant factors in $\Gamma_W$ because it appears in both the numerator and the denominator of Eq.~\eqref{eq:current}.

In order to simplify the numerical calculations, we utilised the symmetries of the system to decompose the numerator of Eq.~\eqref{eq:current} into component tensors, with corresponding numerical coefficients $N_1, N_2, N_3$. The contribution of one cone to the photocurrent is of the form
\begin{multline}
\chi\,v_T^i(N_1 [B^j (v_W^{-2})_{jk} v_T^k] [\epsilon_l (v_W^2)^{lm} \epsilon_m] + \\
N_2 [B^i\epsilon_i] [v_T^j\epsilon_j] +
N_3 [B^j (v_W^{-2})_{jk} v_T^k] [v_T^l\epsilon_l]^2)
\end{multline}
where the scalar coefficients $N_1, N_2, N_3$ depend on $\mu$, $T$, and $\omega$. In general, these coefficients are obtained by numerical integration. 

In a material with tetragonal symmetry, such as the transition metal monopnictides, for light of arbitrary polarization, the magnetic photocurrent has the form
\begin{align}
    \vec{j} =\quad &  A_1[(\hat{\epsilon}\cdot\hat{a})^2 (\hat{\epsilon}\cdot\hat{b})^2] (\hat{c}\times\vec{B})\nonumber\\ + &A_2 (\hat{\epsilon}\cdot\hat{c})^2 (\hat{c}\times\vec{B})\nonumber\\ + &A_3 (\vec{B}\cdot \hat{c})(\hat{\epsilon}\cdot\hat{c})(\hat{\epsilon}\times\hat{c}) \nonumber\\ + &A_4 [\vec{B}\cdot(\hat{\epsilon}\times\hat{c})](\hat{\epsilon}\cdot\hat{c})\hat{c} \nonumber\\ + &A_5 (\hat{\epsilon}\cdot\hat{a})(\hat{\epsilon}\cdot\hat{b})[(\vec{B}\cdot\hat{a})\hat{a} - (\vec{B}\cdot\hat{b})\hat{b}] \nonumber\\ + &A_6 [(\hat{\epsilon}\cdot\hat{a})^2 - (\hat{\epsilon}\cdot\hat{b})^2][(\vec{B}\cdot\hat{a})\hat{b} + (\vec{B}\cdot\hat{b})\hat{a}]. 
\end{align}

We numerically calculated this photocurrent for a prototypical material with tetragonal symmetry. We have assumed 8 tilted Weyl cones ($W_1$) related to each other by reflection and rotation symmetries, and 16 untilted cones ($W_2$), also related to each other by reflection and rotation symmetries, similar to the cone distribution in transition metal monopnictide class of materials which includes TaAs. The energies of $W_1$ cones are 10 meV below the Fermi surface, while the energies of $W_2$ cones are exactly at the Fermi surface (so they contribute to absorption but not to the photocurrent). Both sets of cones have the same untilted Hamiltonians. 

For numerical computations, we have chosen a set of parameters representative of the materials in the TaAs class~\cite{ma2017}. 
The matrix $(v_W^2)^{ij}$ is taken as \begin{equation*}\begin{pmatrix} 16 & 0 & 0\\ 0 & 4 & 0\\ 0 & 0 & 1
\end{pmatrix}\times 10^{10}\; \mathrm{m^2/s^2}. 
\end{equation*}
The tilt velocity for the type-I Weyl semimetal is assumed to be $v_T^i = (1.8, 1.2, 0) \times 10^5\; \mathrm{m/s}$ and for the type-II Weyl semimetal it is assumed to be $v_T^i = (3.6, 2.4, 0) \times 10^5\; \mathrm{m/s}$. These parameters are similar to those in \citet{ma2017}.

We considered light polarized along the $c$ axis, incident on the $ac$ plane of the material, with magnetic field directed along the $b$ axis, so that the observed current would be along the $a$ axis, as predicted by Eq.~\eqref{eq:current1}. The other parameters used in the numerical calculations are given in Table~\ref{table:param} and the results are shown in Figure~\ref{fig:ResultPlot}.
\begin{table}[htb]
\caption{Parameters used for numerical computations}
\centering
\begin{tabular*}{\linewidth}{p{0.2cm} l @{\extracolsep{\fill}} l}
\hline\hline
& Parameter &Value\\ [0.5ex]
\hline
& $\tau$ (relaxation time) &$4 \times 10^{-11}\; \mathrm{s}$ \\
& $B$ (magnetic field) &$0.5 \; \mathrm{T}$ \\
& $R$ (reflectivity) &0.95 \\
& $I$ (light intensity) &$10^{6}\; \mathrm{W/m^2}$\hspace{0.2cm} \\ 
[1ex]
\hline\hline
\end{tabular*}
\label{table:param}
\end{table}
Since the magnitude of the current depends on the tilt $\alpha$, maximal current is predicted for the type-II Weyl semimetals, as shown in Figure~\ref{fig:ResultPlot}. For a beam of frequency $\omega/2\pi = 3\; \mathrm{THz}$ and spot size $0.5\;\mathrm{mm}$ at temperature $T=50\; \mathrm{K}$, the photocurrent is $0.75\; \mathrm{\mu A}$ for the type-I Weyl semimetal and $2.5\; \mathrm{\mu A}$ for the type-II  Weyl semimetal. The photocurrent is obtained by multiplying Eq.~\eqref{eq:current} with the spot size. Note that the resistance of the sample is usually smaller than the load resistance in experimental setups, resulting in a suppression of the detected current.

As seen in Figure~\ref{fig:ResultPlot}, the photocurrent is maximised at around $50\; \mathrm{K}$. The role of temperature is important since transitions cannot happen to completely occupied states; nor can they happen from completely empty states. Temperature allows to avoid the blockade due to Pauli exclusion principle by smearing the occupation fraction. On the other hand, if the temperature is too high, the magnitude of the photocurrent decreases since thermal smearing reduces the difference in occupation fraction between the two states between which transitions can occur.  Note that it is also possible to observe a photocurrent even at zero temperature if the frequency is high enough, since this allows transitions to access the region beyond the sharp Pauli blockade, which can also be seen in Figure~\ref{fig:ResultPlot}.

To summarize, we predict a new type of photocurrent to occur in Weyl semimetals with broken inversion symmetry, time reversal symmetry, and a symmetry with respect to a reflection plane. The class of materials that satisfy these conditions includes the monopnictides such as TaAs. In contrast to all previous proposals and observations of photocurrents in Weyl semimetals, the predicted \emph{transverse chiral magnetic photocurrent} can be induced even by a linearly polarized light and does not require a breaking of reflection symmetry of the crystal; the current is transverse to the direction of an applied magnetic field.  The magnitude of the resulting photocurrent is predicted to be  significant in the THz frequency range, about $0.75\; \mathrm{\mu A}$ for type-I and $2.5\; \mathrm{\mu A}$ for type-II Weyl semimetals. Therefore, the transverse chiral magnetic photocurrent, especially in type-II Weyl semimetals, can enable a significant increase in sensitivity of unpolarized THz radiation detectors. 

\begin{acknowledgments}
We thank Jennifer Cano, Gao Lanlan, Qiang Li and Mengkun Liu for useful and stimulating discussions.
This work was supported in part by the U. S. Department of Energy under Awards DE-SC-0017662 (S. K. and D. E. K.), DE-FG-88ER40388 (E. J. P. and D. E. K.) and DE-AC02-98CH10886 (D. E. K.).
\end{acknowledgments}

\bibliographystyle{apsrev4-1}
\bibliography{bibliography}
\end{document}